\documentclass[12pt]{article}

\usepackage{graphicx}
\usepackage{hyperref}

\begin{document}

\renewcommand{\abstractname}{\hfill}
\baselineskip=20pt

\newpage
\pagenumbering{arabic}
\begin{center}
\large \bf \uppercase{Exact solutions and particle creation for nonconformal
scalar fields in homogeneous isotropic cosmological models}
\end{center}

\begin{center}
\bf
Yu. V. Pavlov${}^*${}\footnote{${}^*$Institute for Problems in Mechanical
Engineering, RAS, St.\,Petersburg, Russia
and
A.\,Friedmann Laboratory for Theoretical Physics, St.\,Petersburg, Russia;
\ E-mail:\, yuri.pavlov@mail.ru}
\end{center}

\begin{abstract}
    The problem is solved of describing scale factors of a homogeneous
isotropic spaces-time such that the exact solution for the scalar field
with a nonconformal coupling to curvature can be obtained from solutions
for the conformally coupled field by redefining the mass and momentum.
    Explicit expressions for dependence of time from the large-scale factor
are presented in the form of Abelian integrals in these cases.
    The exact solution for a scalar field with Gauss-Bonnet type coupling
with curvature is received and it is shown that the corresponding nonconformal
additions can dominate at the particles creation by gravitational field.
\end{abstract}

{\small

}

{\centering \section{\large Introduction}}

    Quantum effects in a curved space-time, in particular, particle creation
by the gravitational field, can have important applications in cosmology and
astrophysics~\cite{GMM},~\cite{BD}.
    Quantum theory in a gravitational field has been studied intensely since
the 1970s.
    The case studied in particular detail is that of conformally  coupled
fields in a homogeneous isotropic space-time.
    Less studied is the case of a nonconformal coupling of a scalar field
to curvature.
    At the same time, nonconformal contributions can dominate both in the
effect of particle creation and in the magnitude of vacuum averages of the
energy-momentum tensor, which determinate the back reaction of a quantized
field on the space-time metric (see, e.g., \cite{BMR},\,\cite{BLMPv}).
    Until recently, no calculation of quantum effects of a scalar field
coupled to a curvature of the Gauss-Bonnet (GB) type have been available.

    Here, we study the possibility of obtaining exact solutions for scalar
fields with both the standard nonconformal and the GB couplings to curvature
from solutions in the case of a conformal coupling in a homogeneous isotropic
space-time.
    We review scaling factors that allow deriving exact solutions for
nonconformal fields.
    We give an exact solution for the scalar field coupled to the GB-type
curvature and analyze particle creation in such a model.
    We calculate particle creation in the quantum field theory in a curved
space-time, where quantized matter fields are considered in an external
classical gravitational field described by the space-time metric.
    The possibility of using such a semiclassical approach relates to a broad
domain between Planckian and Compton characteristic values of length,
curvature, and density~\cite{GMM},~\cite{BD}.

    We use the system of units where $\hbar =c=1$.

\vspace{9mm}
{\centering \section{\large Scalar field in a homogeneous isotropic space}
\label{sec2}}

    We consider a complex scalar field $\varphi(x)$ of mass $m$
with the Lagrangian
    \begin{equation}
L(x)=\sqrt{|g|} \left[\, g^{ik}\partial_i\varphi^*\partial_k\varphi -
(m^2 + V_{\!g})\, \varphi^* \varphi \, \right]
\label{Lag}
\end{equation}
    and the corresponding equation of motion
\begin{equation}
( \nabla^i \nabla_{\! i} + V_{\!g} + m^2 )\, \varphi(x)=0 \,,
\label{Eqm}
\end{equation}
    where ${\nabla}_{\! i}$ are covariant derivatives in the
$N$-dimensional space-time with the metric $g_{ik}$,\
$ g\!=\!{\rm det}(g_{ik})$,  and $\ V_{\!g}$ is a function of invariant
combinations of the metric tensor $g_{ik}$ and its partial derivatives.
    The case $V_{\!g}=0$ corresponds to the minimal coupling of the scalar
field to curvature.
    The case $V_{\!g}=\xi_c R $, where $R$ is the scalar curvature and
$\xi_c = (N-2)/\,[\,4\,(N-1)] $, ia called the conformal coupling to curvature
($\xi_c = 1/6 $ at $N=4$).
    Equation~(\ref{Eqm}) is conformally invariant if $m=0$
and $V_{\!g}=\xi_c R $.

    Taking an arbitrary $ V_{\!g} $ leads to the appearance of third- and
higher-order derivatives of the metric in the metric energy-momentum tensor of
the scalar field and hence in the Einstein equations.
    The appearance of additional higher-derivative terms in the equations,
even if their coefficients are small, leads to a dramatic restructuring of
the theory.
    If we require that the metric energy-momentum tensor not contain
derivatives of the metric of order higher than two, then $ V_{\!g} $
can be chosen as the function
    \begin{equation}
V_{\!g} = \xi R + \zeta R_{GB}^{\,2} \,,
\label{V}
\end{equation}
    where
$$
R_{GB}^{\,2} \stackrel{\rm def}{=}R_{lmpq} R^{\,lmpq} - 4 R_{lm} R^{\,lm} + R^2.
$$
    In the four-dimensional space-time, $R_{GB}^{\,2}$ coincides with the
Euler characteristic density, which by the GB theorem is a topological
invariant (in $N=2$, the corresponding density is proportional to $R$).
    Therefore, the coupling to a curvature of form~(\ref{V}),
introduced in~\cite{Pavlov2004b}, can be naturally called the GB-type coupling.
    In lower dimensions $N=2,3$, new effects due to $\zeta \ne 0$ are absent,
because then $R_{GB}^{\,2}=0$.
    For $N=4$ with constant $\varphi(x)$, the contribution of
the $R_{GB}^{\,2}$ term to the metric energy-momentum tensor is absent because
the corresponding variational derivative vanishes~\cite{Lanczos38}.
    But for a variable $\varphi(x)$, the contribution od such terms must be
taken into account if the constant $\zeta $ of dimension (mass)$^{-2}$ is
nonzero.

    We write the metric of an $N$-dimensional homogeneous isotropic space-time
in the form
\begin{equation}
ds^2 = dt^2 - a^2(t)\, d l^2 = a^2(\eta)\,(d{\eta}^2 - d l^2) \,,
\label{gik}
\end{equation}
    where $d l^2=\gamma_{\alpha \beta} d x^\alpha d x^\beta $ is the metric
of the $(N-1)$-dimensional space of constant curvature $K=0, \pm 1 $.
    The full system of solutions of Eq.~(\ref{Eqm}) in metric~(\ref{gik})
can be found in the form
    \begin{equation}
\varphi(x) = a^{-(N-2)/2} (\eta)\, g_\lambda (\eta) \Phi_J ({\bf x}) \,,
\label{fgf}
\end{equation}
    where
    \begin{equation}
g_\lambda''(\eta)+\Omega^2(\eta)\,g_\lambda(\eta)=0 \,,
\label{gdd}
\end{equation}
       \begin{equation}
\Omega^2(\eta)=m^2 a^2 +\lambda^2 - \Delta \xi\,a^2 R +
\zeta a^2 R_{GB}^{\,2} \,,
\label{Ome}
\end{equation}
     \begin{equation}
\Delta_{N-1}\,\Phi_J ({\bf x}) = - \Biggl( \lambda^2 -
\biggl( \frac{N-2}{2} \biggr)^2 K \Biggr) \Phi_J  ({\bf x})\,,
\label{DFlF}
\end{equation}
    $\Delta \xi =\xi_c - \xi $,
the prime denotes the derivative with respect to the conformal time~$\eta$,
and $J$ is the set of indices (quantum numbers) that label eigenfunctions of
the Laplace-Beltrami operator $\Delta_{N-1}$ in the ($N-1$)-dimensional space.
    In metric~(\ref{gik}), the expressions for the scalar curvature and the
GB invariant are~\cite{Pavlov2004b}
     \begin{equation}
R= a^{-2}(N-1) \left[ \, 2c'+(N-2)(c^2+K) \right],
\label{RRRR}
\end{equation}
       \begin{equation}
R_{GB}^{\,2} = a^{-4} (N\!-\!1) (N\!-\!2) (N\!-\!3) (c^2 +K)
\left[\, 4 c'+ (N\!-\!4) \, (c^2 +K) \,\right],
\label{RGBoi}
\end{equation}
    where $c=a'/a = \dot{a}(t)$.

    In accordance with the method of diagonalization of
a Hamiltonian~\cite{GMM} (see~\cite{PvIJA} for the case of an
arbitrary $V_{\!g}$), the functions $g_\lambda(\eta)$ must satisfy the initial
conditions
    \begin{equation}
g_\lambda'(\eta_0)=i\, \Omega(\eta_0)\, g_\lambda(\eta_0)\,, \ \ \ \ \
|g_\lambda(\eta_0)|= \Omega^{-1/2}(\eta_0)\,.
\label{icg}
\end{equation}
    If the quantized scalar field is in the vacuum state at
the instant~$\eta_0$, then the number density of the pairs of particles
created up to the instant~$\eta $ can be evaluated (for $K=0$) as~\cite{GMM}
    \begin{equation}
n(\eta) = \frac{B_N}{2 a^{N-1}} \int \limits_0^\infty \! S_\lambda(\eta)\,
\lambda^{N-2}\, d \lambda,
\label{nN}
\end{equation}
where $B_N=\left[2^{N\!-3} \pi^{(N \!-1)/2} \Gamma((N\!-1)/2) \right]^{-1}\!,$
\ $\Gamma(z)$ is the gamma function, and
    \begin{equation}
S_\lambda(\eta) = \frac{\left| g'_\lambda (\eta ) -
i \Omega \, g_\lambda (\eta ) \right|^2}{4 \Omega } \,.
\label{Sgg}
\end{equation}
    As shown in~\cite{Pavlov2001}, $ S_\lambda \sim \lambda^{-6} $,
and the integral in~(\ref{nN}) converges for $N<7$.

\vspace{9mm}
{\centering \section{\large On exact solutions for different types of
coupling to curvature}
\label{secTochR}}

    With the substitution $ g(\eta) = \exp z(\eta) $, Eq.~(\ref{gdd})
reduces to the Riccati equation of the general type
for $v(\eta) \equiv z'(\eta) $:
    \begin{equation}
v'(\eta)+ v^2(\eta) + \Omega^2(\eta)=0 ,
\label{Riccati}
\end{equation}
    Therefore, the number of scaling factors $a(\eta)$ admitting exact
solutions is relatively small.
    In the cases where an exact solution can nevertheless be found, it is
typically expressed in terms of special functions:
hypergeometric, Bessel functions, etc.
    Several exact solutions are given in~\cite{GMM}, \cite{BD}.
    Scaling factors admitting exact solutions are briefly reviewed
in~\cite{PavlovGr702009},~\cite{PavlovUZKU}.

    In a homogeneous isotropic space-time, Eq.~(\ref{gdd}) has a simpler form
for the field with the conformal coupling\, $\Omega^2 = m^2 a^2 + \lambda^2$\,
than for a nonconformal coupling.
    This does not forbid the existence of exact analytic solutions of a more
complicated equation for a nonconformal scalar field.
    But the majority of the known exact solutions for a nonconformal scalar
field are obtained from solutions for the conformal field by replacing the
mass and momentum parameters.

    We consider the following question.
    At what scaling factors can the exact solution for the scalar field with
a nonconformal coupling to curvature be obtained from a solution for the
field with conformal coupling by redefining the values of the field mass~$m$
and the dimensionless momentum~$\lambda$?

    For the scalar field with a nonconformal coupling of the form~$\xi R$,
in accordance with~(\ref{Ome}), this is to be the case
if there exist constants~$\beta_1$ and $\beta_2$ such that
    \begin{equation}
\Delta \xi R\, a^2 = \beta_1 + \beta_2 a^2 .
\label{bet1}
\end{equation}
    This condition, with~(\ref{RRRR}) taken into account, is an ordinary
second-order differential equation for $a(\eta)$, which reduces to
a linear first-order equation for $ f(a) = a^{\, \prime\, 2}(\eta)$.
    Its solution can be represented in the form
    \begin{equation}
\int \frac{d a}{\sqrt{C_0 a^{4-N} +C_1 a^2 + C_2 a^4}} = \pm (\eta - \eta_0),
\label{bet2}
\end{equation}
    where $C_0, C_1, C_2$, and $\eta_0$ are arbitrary real constants such that
$C_0 a^{4-N} +C_1 a^2 + C_2 a^4 > 0 $.

    In the general case for $N>4$, integral~(\ref{bet2})
is hyperelliptic, and calculating $a(\eta)$ amounts to inverting
the hyperelliptic integral~\cite{MarkushevichVKTAF}.
    For $N=4$, obtaining $a(\eta)$ from (\ref{bet2})
is the problem of inversion an elliptic integral~\cite{BatemanErdelyi}.
    In particular cases where integral~(\ref{bet2}) is pseudoelliptic
or $N=4$ and $C_2=0$, it can be expressed in elementary functions.
    We give examples of such scaling factors.

    The first example of a scaling factor is given by the relations
    \begin{equation}
C_1=C_2=0 , \ \ \ \ \
a= a_1 \eta^{2 / (N-2)} = a_0 t^{2/N} , \ \ \ \ a^2 R = (N-1)(N-2) K ,
\label{bet1p}
\end{equation}
    where $ a_1, a_0 = {\rm const}$. \
    For $N=4$ with $K=0$, such a scaling factor corresponds to a case
extremely important from the application standpoint, the radiation-dominated
universe.
    Solutions are known for $N=4$, where they can be expressed in terms of
Kummer's hypergeometric function, and for $N=6$, where they can be expressed
in terms of Bessel functions~\cite{PavlovGr702009},~\cite{MMStarobinsky}.

    The second example of a scaling factor is
    \begin{equation}
C_0=C_2=0 , \  \ \ \ a= a_1 e^{\alpha \eta} = \alpha t  , \
\ \ \ \ a^2 R = (N-1)(N-2)(\alpha^2 + K) ,
\label{bet2p}
\end{equation}
    where $ \alpha = {\rm const}$. \
    For $ \alpha =1 , K=-1$, this is Milne's universe~\cite{Milne32}
(i.e., a part of Minkowski space in the appropriate coordinates).
    A solution of Eq.~(\ref{gdd}) expressed in terms of a Bessel function
was presented in~\cite{PavlovGr702009}.

    The third example of a scaling factor is
    \begin{equation}
C_0=C_1 =0 , \ C_2 = H^2, \ \ \
a= \frac{-1}{H \eta} = \frac{e^{H t}}{H}  , \ \ \ \
a^2 R = (N\!-\!1) \left[ N H^2 a^2 + (N \!-\! 2) K \right]\!,
\label{bet3p}
\end{equation}
    where $ H = {\rm const}$. \
    For $ K=0$, corresponding coordinates describe a part of
the De Sitter universe.
    The solution of Eq.~(\ref{gdd}) can be expressed in terms of a Hankel
function~\cite{GMM}.

    For $C_0 =0 $, solutions of Eq.~(\ref{bet2}) are also given by
    \begin{equation}
\frac{a_1}{\cosh \gamma \eta}  = a_1 \sin \frac{\gamma t}{a_1}, \ \ \ \
\frac{-a_1}{\sinh \gamma \eta}  = a_1 \sinh \frac{\gamma t}{a_1}, \ \ \ \
\frac{a_1}{\cos \gamma \eta}  = a_1 \cosh \frac{\gamma t}{a_1}\,,
\label{sch4xi}
\end{equation}
    where $ \gamma = {\rm const}$. \
    In the four-dimensional space-time, solutions of Eq.~(\ref{bet2})
with $C_1^2 =4 C_0 C_2$ are
    \begin{equation}
    \begin{array}{c}  \displaystyle
a_1 \tan \gamma \eta  = a_1 \sqrt{\exp \frac{2 \gamma t}{a_1} - 1 },
\ \ \ \ \
a_1 \tanh \gamma \eta  = a_1 \sqrt{1 - \exp \frac{-2 \gamma t}{a_1} },
    \\[11pt]   \displaystyle
- a_1 \coth \gamma \eta  = a_1 \sqrt{\exp \frac{2 \gamma t}{a_1} + 1 },
\hfill
\end{array}
\label{tg4xi}
\end{equation}
    and solutions with $C_2 =0$ are
    \begin{equation}
a_1 \sin \gamma \eta  = \sqrt{a_1^2 \!-\! \gamma^2 t^2},
\ \ \ \ a_1 \sinh \gamma \eta  = \sqrt{\gamma^2 t^2 - a_1^2},
\ \ \ \ a_1 \cosh \gamma \eta  = \sqrt{\gamma^2 t^2 + a_1^2},
\label{bet4}
\end{equation}
    Solutions of Eq.~(\ref{gdd}) with scaling factors~(\ref{sch4xi}) and
(\ref{tg4xi}) can be expressed in terms of the hypergeometric functions.
    Equation~(\ref{gdd}) with scaling factor~(\ref{bet4}) reduces to
a (modified) Mathieu equation.

    We note that if a solution of Eq.~(\ref{gdd}) with a scaling factor
$a(\eta)$ is found, then it can be used to obtain a solution for the scaling
factor $\tilde{a} = \sqrt{a^2(\eta) + b^2} $, where $ b = {\rm const} $,
by redefining~$\lambda$.

    With the coordinate time $t$, solutions of Eq.~(\ref{bet1}) can be
represented in the form
    \begin{equation}
\int \frac{d \left( a^{N/2} \right)}{\sqrt{C_0 + C_1 a^{\mathstrut N-2} +
C_2 a^N}} = \pm \frac{N}{2} (t - t_0),
\label{bet2t}
\end{equation}
    where $t_0 = {\rm const}$. \
    For $N=4$, integral~(\ref{bet2t}) is expressed in elementary functions.
    Therefore,
{\it in the four-dimensional space-time, the dependence of the coordinate
time~$t$ on scaling factor under which solutions for the scalar field with
a $\xi R$ coupling to curvature can be obtained from solutions for the field
with conformal coupling by redefining the mass and momentum values must be
expressible in a finite form in terms of elementary functions.}

    We consider the case of a scalar field with the GB coupling to the
curvature in~(\ref{V}) and assume that $N\ge 4$ and $\zeta \ne 0$.
    An exact solution of Eq.~(\ref{gdd}) can be obtained from the solution
for the conformally coupled field by redefining $m$ and $\lambda$ if there
exist constants~$\tilde{\beta}_1$ and $ \tilde{\beta}_2$ such that
    \begin{equation}
-\Delta \xi R\, a^2 + \zeta R_{GB}^{\,2} a^2 =
\tilde{\beta}_1 + \tilde{\beta}_2 a^2 .
\label{bgb1}
\end{equation}
    With formulas~(\ref{RRRR}) and (\ref{RGBoi}), this condition reduces to
a first-order linear inhomogeneous differential equation for the function
    \begin{equation}
F(a) = \left(
\dot{a}^2(t) +K - \frac{\Delta \xi \, a^2}{2 \zeta (N-2) (N-3) } \right)^2 .
\label{DUF}
\end{equation}
    Solving differential equation~(\ref{bgb1}) for $a(t)$, we obtain an
expression for the dependence of the scaling factor on the coordinate time:
    \begin{equation}
\int \frac{d a}{\sqrt{ \displaystyle -K +
\frac{\Delta \xi \, a^{\mathstrut 2}}{2 \zeta (N-2)(N-3) }
\pm \sqrt{D_0 a^{\mathstrut 4-N} +D_1 a^2 + D_2 a^4}}} = \pm (t - t_0),
\label{bgb2}
\end{equation}
    where $D_0, D_1$, and $D_2$ are arbitrary real constants such that the
integral is real.

    In the general case, finding the explicit form of $a(t)$ amounts
to inverting Abelian integral~(\ref{bgb2}).
    We give examples in particular cases:
    \begin{equation}
K = \Delta \xi = D_1 = D_2 = 0 , \ \ \ \
a= \gamma t^{4/N} , \ \ \  R_{GB}^{\,2} =0,
\label{pGB1}
\end{equation}
    where exact solutions for the conformal coupling are known for
$N=4,8,12$ (see formulas~(\ref{bet1p}) and (\ref{bet2p}));
    \begin{equation} \label{pGB2}
K = D_0 = D_1 = 0 ,  \ \ \ \ a= a_0 \exp H t ,
\ \ \ \ R_{GB}^{\,2} = N(N\!-\!1)(N\!-\!2)(N\!-\!3)H^4
\end{equation}
    (see formulas~(\ref{bet3p}));
and for $N=4$, functions~(\ref{sch4xi}) are solutions of Eq.~(\ref{bgb2})
with $D_1^2 = 4 D_0 D_2$.

    In what follows, we consider the exact solution and particle creation
for the first scaling factor in~(\ref{sch4xi}).

\vspace{9mm}
{\centering \section{\Large \large  Particle creation in the
$ a_1 \sin (\gamma t / a_1) = a_1 / \cosh \gamma \eta $ model}
\label{secch}}

    Space-time with a scaling factor
$ a_1 \sin (\gamma t / a_1) = a_1 / \cosh \gamma \eta $\,
evolves between two singularities at $t=0 \ \, (\eta = -\infty)$
and $t=T \equiv \pi  a_1 / \gamma \ (\eta = +\infty)$,
if $\gamma^2+K \ne 0$, \, $N\ge3$.
    In the case $\gamma =1$ and $K=-1$, the corresponding coordinates cover
a part of the De Sitter space of the second kind~\cite{HawkingEllis}
with a constant negative curvature $R=-(N-1)N/a_1^2$.

    Particle creation in this model was previously considered for a conformal
coupling to the curvature in~\cite{GMM}, and the case of the $\xi R $
coupling was investigated in~\cite{BMR}.
    We here consider the creation of scalar particles with a GB-type
coupling to the curvature, Eq.~(\ref{V}), in a four-dimensional space-time.
    From~(\ref{RRRR}) and (\ref{RGBoi}) with $N=4$, we obtain
    \begin{equation}
R a^2 = 6 \left( \gamma^2 +K \right) - \frac {12 \gamma^2}{a_1^2}\, a^2,
\ \ \ \ a^2 R_{GB}^{\,2} = -\frac{24 \gamma^2}{a_1^2} \left[
K + \gamma^2 \left(1 - \frac{a^2}{a_1^2} \right) \right],
\label{RGB3}
\end{equation}
    and therefore Eq.~(\ref{gdd}) becomes
    \begin{equation}
g''(\eta) + \left[ \left( \! m^2 a_1^2 + \Delta \xi 12 \gamma^2
+ \zeta \frac{24 \gamma^4 }{a_1^2} \right)\! \frac{1}{\cosh^2 \! \gamma \eta} +
\lambda^2 - 6 (\gamma^2 + K) \! \left( \! \Delta \xi +
\zeta \frac{4 \gamma^2}{a_1^2} \right) \right] g(\eta) = 0.
\label{gch}
\end{equation}
    The solution of Eq.~(\ref{gch}) with initial conditions~(\ref{icg})
with $\eta \to - \infty$ can be obtained from the exact solution for
the conformally coupled case~\cite{GMM} by redefining the values of the mass
($m \to M$) and momentum ($\lambda \to \Lambda$):
    \begin{equation}
M^2 =  m^2 + \frac{12 \gamma^2}{a_1^2}
\left( \Delta \xi + \zeta \frac{2 \gamma^2}{a_1^2} \right), \ \ \ \
\Lambda^2 = \lambda^2 - 6  ( \gamma^2 + K )
\left( \Delta \xi + \zeta \frac{4 \gamma^2}{a_1^2} \right).
\label{mltilde}
\end{equation}
    The solution has the form
    \begin{equation}
g(\eta) = \frac{e^{i (\Lambda \eta + \alpha_0)}}{\sqrt{\Lambda}}
F \biggl( A, B;\, C;\, \frac{1+ \tanh \gamma \eta}{2} \biggr),
\label{getch}
\end{equation}
    where $F(A, B; C; z)$ is the hypergeometric function,
$ \alpha_0$ is an arbitrary real constant, and
    \begin{equation}
A= \frac{1}{2} + \sqrt{ \frac{1}{4} + \frac{M^2 a_1^2}{\gamma^2} },
\ \ \ \
B= \frac{1}{2} - \sqrt{ \frac{1}{4} + \frac{M^2 a_1^2}{\gamma^2} },
\ \ \ \
C=1 + i \frac{\Lambda}{\gamma}.
\label{ABC}
\end{equation}

    The condition $\Omega^2 \ge 0$, which is necessary in the Hamiltonian
diagonalization method, is satisfied for Eq.~(\ref{gch})
for any~$\lambda$ and $\eta$ if
    \begin{equation}
( \gamma^2 + K ) \left( \Delta \xi + \zeta \frac{4 \gamma^2}{a_1^2} \right)
\le 0, \ \ \ \ \
m^2 a_1^2 + 6 (\gamma^2 -K) \Delta \xi - 24 \zeta K \frac{\gamma^2}{a_1^2}
\ge 0 .
\label{nom0}
\end{equation}
    In particular,\, $\Omega^2 \ge 0$, \ at $K=0$ and
    \begin{equation}
- \zeta \frac{4 \gamma^2}{a_1^2} \ge \Delta \xi \ge
- \frac{m^2 a_1^2}{6 \gamma^2} \,.
\label{nxzetK0}
\end{equation}

    From asymptotic expression of~(\ref{Sgg}) for exact solution~(\ref{getch}),
similarly to the conformal case~\cite{GMM}, we obtain
the limit spectrum (as $\eta \to +\infty$) of the created particles:
    \begin{equation}
S_\lambda = \frac{\displaystyle \cos^2 \frac{\pi}{2} \sqrt{1 +
\frac{4 M^2 a_1^2}{\gamma^2}} }{\displaystyle
\sinh^2 \frac{\pi \Lambda}{ \gamma} }.
\label{Slim}
\end{equation}
    In accordance with formulas~(\ref{nN}), (\ref{mltilde}), and (\ref{Slim}),
the number of particle pairs created in a Lagrangian volume $a^3(t)$ at $K=0$
per the evolution cycle  $ t \to T$ is given by
    \begin{equation}
N= \frac{\gamma^3 }{2 \pi^5} \cos^2 \pi \sqrt{\frac{1}{4} +
\frac{m^2 a_1^2}{\gamma^2} + 12 \Delta \xi + 24 \zeta \frac{\gamma^2}{a_1^2}}
\ \int \limits^\infty_{x_0} \frac{ x \sqrt{x^2- x_0^2}}{\sinh^2 x} \, dx ,
\label{nlx}
\end{equation}
    where
    \begin{equation}
x_0 = \pi \sqrt{-6\, \Bigl( \Delta \xi + \zeta \frac{4 \gamma^2}{a_1^2}
\Bigr)}.
\label{nlx0}
\end{equation}
    Because~\cite{GradRizh}
    \begin{equation}
\int \limits_0^\infty \frac{x^2}{\sinh^2 x}\, d x = \frac{\pi^2}{6} ,
\label{intgr}
\end{equation}
    it follows from formulas~(\ref{mltilde}), (\ref{nxzetK0}), and (\ref{nlx0})
that
    \begin{equation}
\Delta \xi = - \zeta \frac{4 \gamma^2}{a_1^2} \ge
- \frac{m^2 a_1^2}{6 \gamma^2} \ \ \ \Rightarrow \ \ \
N= \frac{\gamma^3 }{12 \pi^3} \cos^2 \frac{\pi}{2} \sqrt{1  + 24 \Delta \xi +
\frac{4 m^2 a_1^2}{\gamma^2}} .
\label{nlxdd}
\end{equation}

    Formulas~(\ref{Slim}) and (\ref{nlx}) reproduce the corresponding results
for the scalar field with the $\xi R$ coupling at $\zeta =0$ \cite{BMR} and
for conformally coupled particles at $\zeta =0 = \Delta \xi$~\cite{GMM}.
    At $\Delta \xi =0 $, formula~(\ref{nlxdd}) gives the number of pairs
of conformally coupled particles created per the evolution cycle:
    \begin{equation}
N_0 = \frac{\gamma^3 }{12 \pi^3} \cos^2 \frac{\pi}{2} \sqrt{1 +
\frac{4 m^2 a_1^2}{\gamma^2}} .
\label{nlxconf}
\end{equation}

\begin{figure}[ht]
\centering
  \includegraphics[width=9cm]{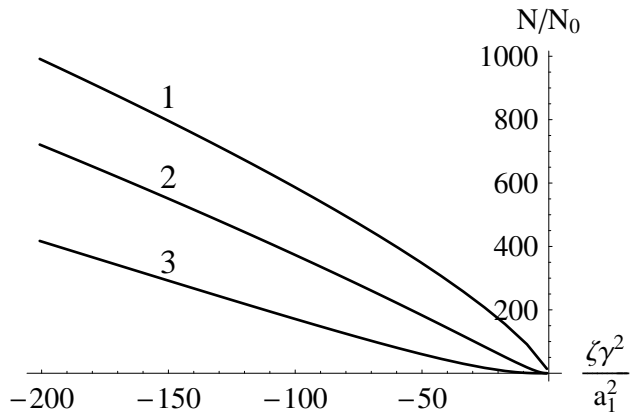}
\caption{Dependence of the number of created particles on
the parameter~$\zeta$: \newline
curve~1, $\Delta \xi = -1/6 $; \, curve~2, $\Delta \xi = 1 $; \,
and curve~3, $\Delta \xi = 3 $.}
\label{figGB}
\end{figure}

    If $\zeta \to -\infty$ for a fixed $\Delta \xi$, then $ M^2 \to - \infty $,
and the number of created particles increases indefinitely, as is shown,
for example, for $m a_1/ \gamma=1$ in Fig.~\ref{figGB}.

    It follows from~(\ref{Slim}) that particle creation per evolution cycle
is absent if
    \begin{equation}
M^2 a_1^2 = n(n+1) \gamma^2, \ \ \ \ n = 0, 1 , 2 , \ldots
\label{nlim}
\end{equation}
    With expression~(\ref{mltilde}) and inequalities~(\ref{nxzetK0}) with $K=0$
taken into account, there is no particle creation per evolution cycle if the
parameter~$\zeta$ takes the values
    \begin{equation}
\zeta_n = \frac{a_1^2}{24 \gamma^2} \left( n(n+1) - 12 \Delta \xi -
\frac{m^2 a_1^2}{ \gamma^2} \right)  , \ \ \ \ n = 0, 1 , \ldots \le
\sqrt{\frac{m^2 a_1^2}{ \gamma^2} + 6 \Delta \xi + \frac{1}{4} } - \frac{1}{2}.
\label{nlim2}
\end{equation}

    Hence, depending on the values of $\zeta$, particle creation in this
model can either increase without bound or be absent altogether.
    Therefore, the effect of the parameter~$\zeta$ of the scalar field
coupling to the GB invariant can dominate, and it must be taken into
account in calculating of the effects of scalar fields in a curved
space-time.

\vspace{7mm}
 {\bf Acknowledgments.}
    The author thanks Professor A.\,A. Grib and the participants in the seminar
at the A.\,Friedmann Laboratory for Theoretical physics for the discussion
of this work.
    The research is done in collaboration with Copernicus Center for
Interdisciplinary Studies, Krak\'{o}w, Poland and supported by the grant from
The John Templeton Foundation.

\vspace{9mm}

\end{document}